\documentclass[letterpaper]{jpconf}
\usepackage{graphicx}

\def\bz{B^0}
\def\bzkx{\bz \to K^{*0} X^0}
\def\bzkxmu{B^0 \to K^{*0} X^0,~K^{*0}\to K^+ \pi^-,~X^0 \to \mu^+ \mu^-}

\def\bzrhox{\bz \to \rho^{0} X^0}
\def\bzrhoxmu{\bz \to \rho^{0} X^0,~\rho^{0} \to \pi^+ \pi^-,~X^0 \to \mu^+ \mu^-}

\newcommand{\mbc}{{M_{\textrm{bc}}}}

\newcommand{\BR}{{\mathcal B}}

\newcommand{\gev}{{\hbox{GeV}}}

\newcommand{\mgev}{{\hbox{GeV}/c^2}}

\begin{document}
\title{Search for an exotic light particle at Belle}

\author{HyoJung Hyun (on behalf of the Belle collaboration)}
\address{Kyungpook National University, Daegu 701-702, Korea}
\ead{hjhyun@gmail.com}

\begin{abstract}
The HyperCP experiment reported the observation of three events for the $\Sigma^+ \rightarrow p \mu^+\mu^-$
decay. The dimuon masses of the observed events are clustered within the detector
resolution of 1 MeV/$c^2$. These decays might be interpreted as a two-body decay, $\Sigma^+ \rightarrow pX^0$, 
$X^0 \rightarrow \mu^+\mu^-$, where $X^0$ is a new particle with mass (214.3 $\pm$ 0.5) MeV/$c^2$.
We report on a search for the $X^0$ particle in $\bzkx$ and $\bzrhox$ decays 
using 656 million $B$ meson pairs collected with the Belle detector at the KEKB asymmetric energy
$e^+e^-$ collider. 
\end{abstract}

\section{Introduction}
The HyperCP experiment~\cite{HyperCP} has reported the observation of three events for $\Sigma^+ \rightarrow p \mu^+ \mu^-$ decay. 
The dimuon masses of the observed events are clustered within the detector resolution of 1 MeV/$c^2$, around 214.3 MeV/$c^2$. 
These decays might be interpreted as a two-body decay, $\Sigma^+ \rightarrow p X^0,~X^0 \rightarrow \mu^+ \mu^-$. 
The probability for all three events to be anywhere in kinematically allowed region within the detector resolution is less than 1\% within the standard model (SM).
An existence of a new particle with such a low mass would be a signal for a new physics beyond the SM.
There are particularly intriguing interpretations for the $X^0$; a pseudoscalar sgoldstino particle~\cite{sgoldstino} 
in various supersymmetric models~\cite{susy},
a light pseudoscalar Higgs boson~\cite{higgs} in the Next-to-Minimal-Supersymmetric Standard Model as well as a vector $U$-boson~\cite{uboson}. The result of model independent study for a spin-1 $X^0$ particle is also available in Ref.~\cite{oh}.
In particular, the estimated branching fractions for $B \to VX^0,~X^0 \to \mu^+\mu^-$ where $X^0$ is a sgoldstino particle 
and $V$ is either a $K^{*0}$ or $\rho^0$ mesons, are in the range $10^{-9}$ to $10^{-6}~\cite{demidov}$.
We report on a search for $X^0$ decaying to dimuon with the decay modes, $\bzkxmu$ and $\bzrhoxmu$ 
using a data sample of $657 \times 10^6~B \bar B$ pairs collected with the Belle detector~\cite{belled}
at the KEKB asymmetric-energy $e^+ e^-$ collider~\cite{KEKB}.

\section{KEKB and Belle detector}
The KEK B-Factory, KEKB, is an asymmetric $e^+e^-$ collider operating at the $\Upsilon$(4S).
Beam energies are 3.5 GeV for Low Energy Ring (LER, $e^+$) and 8 GeV for High Energy Ring (HER, $e^-$).
The beams collide at a single interaction point with a crossing angle of $\pm$11 mrad.
The mass of the $\Upsilon$(4S) is just above the threshold of $B \bar B$ production and exclusively decays to $B^0\bar{B^0}$ or $B^+B^-$.

The Belle detector is a large-solid-angle magnetic spectrometer that consists of a silicon vertex detector (SVD),
a 50-layer central drift chamber (CDC), an array of aerogel threshold Cherenkov counters (ACC),
a barrel-like arrangement of time-of-flight scintillation counters (TOF), and an electromagnetic calorimeter (ECL)
comprised of CsI(Tl) crystals located inside a superconducting solenoid coil that provides a 1.5~T
magnetic field. An iron flux-return located outside of the coil is instrumented to detect $K_L^0$ mesons and to identify
muons (KLM).
The muon detection efficiency and fake rate of KLM are better than 90\% and less than 5\%, respectively.

\section{Search for an exotic light particle}
In this search, we assume that the light $X^0$ particle is a scalar particle and decays to dimuon immediately with $\BR(X^0 \to \mu^+ \mu^-)$. 
Unless specified otherwise, charge-conjugate modes are implied. 
The term scalar $X^0$ particle implies either a scalar or pseudoscalar particle throughout this proceeding.. 

The reconstruction of $K^{*0}$ ($\rho^0$) in the $\bzkx$ ($\bzrhox$) decay uses identified $K^+$ ($\pi^+$) and $\pi^-$ ($\pi^-$) tracks. 
The invariant mass of the $K^{*0}$ ($\rho^0$) candidate is required to be within $\pm$1.5$\Gamma$ ($\pm$1.0$\Gamma$) of the central value.
The $\mu^+\mu^-$ tracks are used to reconstruct $X^0$ candidates.
$B^0$ candidates are reconstructed from a $K^{*0}$ ($\rho^0$) candidate and a $\mu^+\mu^-$ pair. 
Reconstructed $B^0$ candidates are selected using kinematic variables, such as the beam energy-constrained mass $M_{bc}$ and energy difference $\Delta E$.
$M_{bc}$ and $\Delta E$ are defined as follows; 
$M_{bc} = \sqrt{E_\textrm{beam}^2-p_{B}^2}$ and $\Delta E = E_B - E_\textrm{beam}$,
where $E_\textrm{beam}$ is the beam energy and $E_B$ ($p_B$) are the energy (momentum) of the reconstructed $\bz$ candidates
evaluated in the center-of-mass frame.
In the events containing more than one $B^0$ candidate, we select a $B^0$ candidate having the smallest $\chi^2$ value, 
where the $\chi^2$ value is calculated by one-vertex constrained fit using four charged tracks in the candidates.
The signal region for dimuon mass of HyperCP event search is defined to be  
211.6 MeV/$c^2$ $<$ $M_{\mu^+\mu^-}$ $<$ 217.2 MeV/$c^2$ by considering the uncertainties
in $X^0$ mass measurement of the HyperCP and Belle detector resolutions.
The signal efficiency is estimated by accepted events in the $X^0$ mass window for the selected events 
in the $\Delta E$ - $\mbc$ signal region for each mode. 

For background studies, Monte Carlo (MC) samples which are three times larger than the data sample are used.
The MC sample is fitted with a probability density function, $(x-0.21)^n$ for $x$ $>$ 2$m_{\mu}$, where $x$ is a dimuon mass in GeV/$c^2$, 
$m_{\mu}$ is the muon mass and the parameter $n$ is extracted from the fit at sideband region. 
The sideband region is defined as $-0.12$ $\gev$ $<$ $\Delta E$ $<$ $-0.06$ $\gev$ and 0.06 $\gev$ $<$ $\Delta E$ $<$ 0.12 $\gev$, and
5.25 $\mgev$ $<$ $\mbc$ $<$ 5.27 $\mgev$.
The estimated backgrounds for the HyperCP event search are 0.13$^{+0.04}_{-0.03}$ and 0.11$^{+0.03}_{-0.02}$ 
for $\bzkx$ and $\bzrhox$, respectively.

\begin{figure}[t]
\centering
\includegraphics[width=0.4\textwidth]{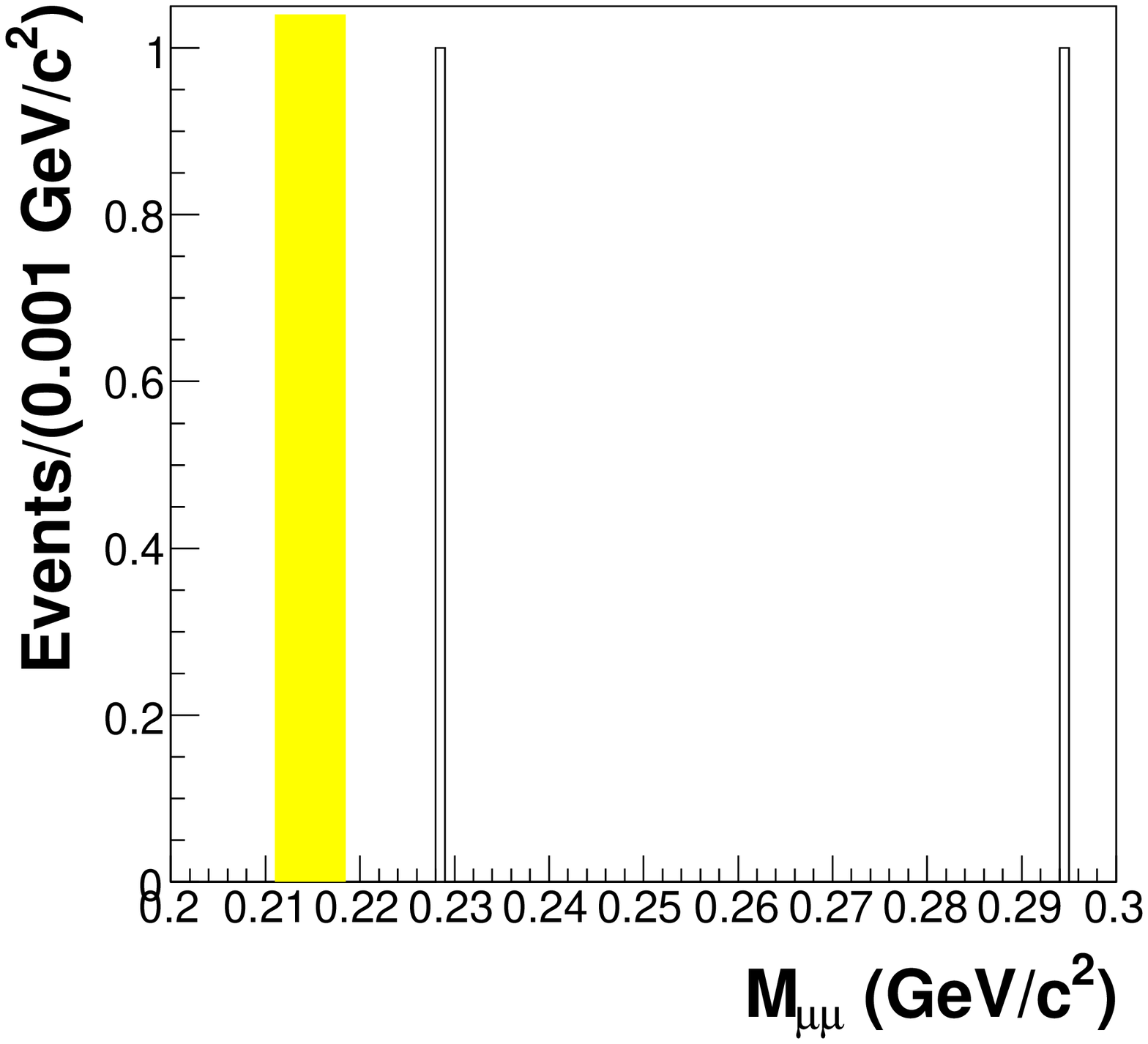}
\includegraphics[width=0.4\textwidth]{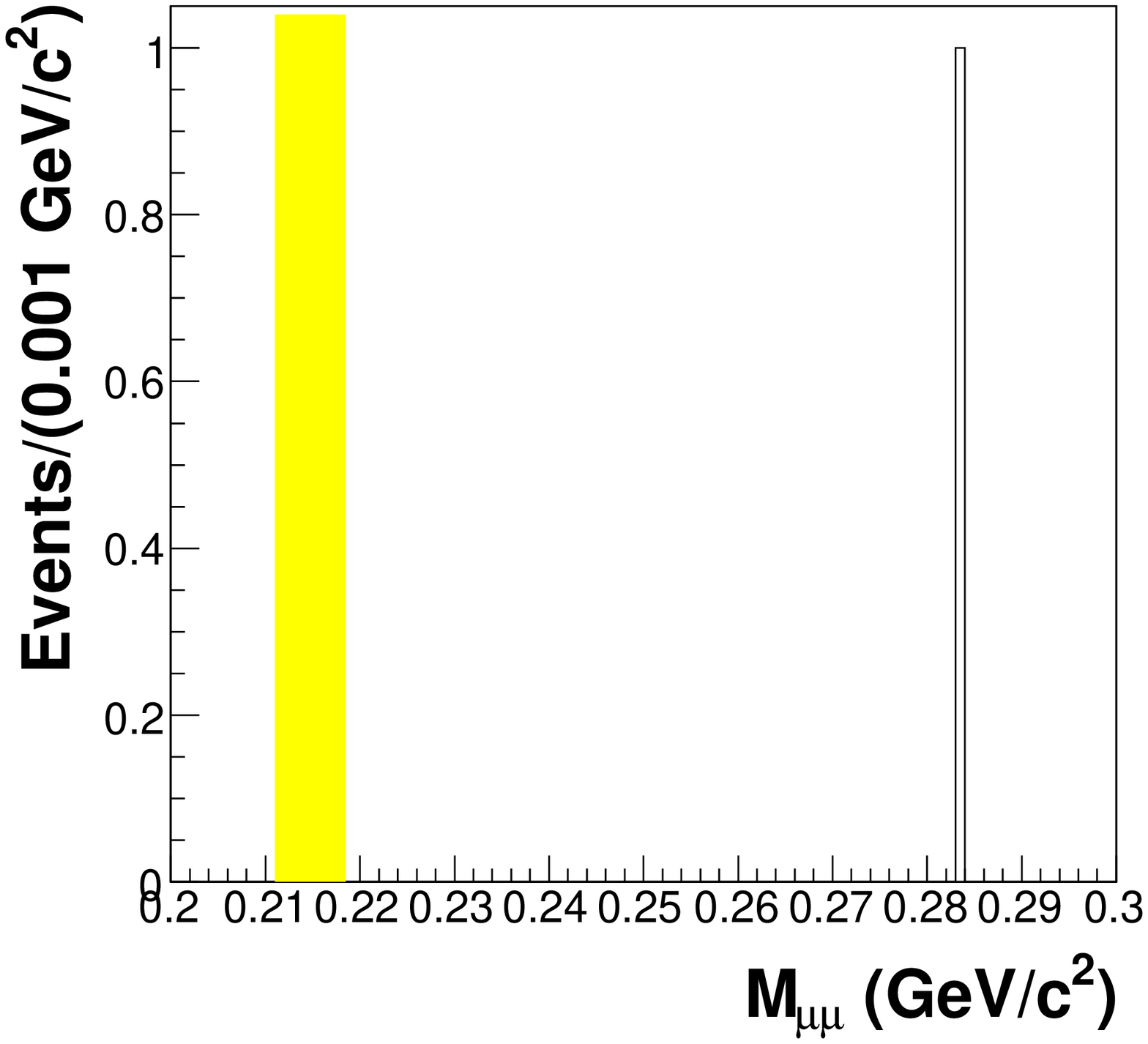}\\
\centering
\caption{The results for dimuon mass distributions for $\bzkx$ (left) and $\bzrhox$ (right). 
	Shaded region in each plot represents for the signal region of the $X^{0}$ search.}
\label{fig_OpenBox}
\end{figure}

Before examining the all data sample, we compare various distributions such as $\mbc$, $\Delta E$, dimuon mass, and etc. with a part of the MC and data samples.
These are in good agreement. For the full data sample, there is no significant signal for $\bzkx$ and $\bzrhox$ decays for the inspected dimuon mass region.  
Figure~\ref{fig_OpenBox} shows the dimuon mass distributions in the signal region for HyperCP event search which is represented as shaded region.

The total systematic uncertainties in the upper limits for the decays $\bzkx$ and $\bzrhox$ are 5.2\% and 5.7\%, respectively.
The dominant systematic uncertainties come from tracking efficiency and muon identification.
The uncertainty for the tracking efficiency is estimated by linearly summing the single track systematic errors which are about 1\%/track.
The systematic for the muon identification is measured as a function of momentum and direction by using the $\gamma\gamma \to \mu^+\mu^-$ data sample.

The upper limits at 90\% C.L. are calculated with the following formula. 

\begin{eqnarray}
\BR (\bz \to V X, X \to \mu^+ \mu^-)< \frac{S_{90}}{\epsilon \times N_{B \bar B} \times \mathcal{B}_{V}}, \nonumber
\label{eq:br}
\end{eqnarray}
 
\hspace{-0.2in}where $V$ stands for either $K^{*0}$ or $\rho^0$, and $\mathcal{B}_V$~\cite{pdg} are the intermediate vector meson branching fractions,
$\BR(K^{*0} \to K^+ \pi^-$) or $\BR(\rho^0 \to \pi^+ \pi^-)$.
Here $N_{B \bar B}$ and $\epsilon$ denote the number of $B \bar B$ pairs and the signal efficiency with small data/MC corrections for
charged particle identification, respectively.
$S_{90}$ is signal yield which is obtained from hypothetical experiment with the number of observed events, the number of background, and systematic uncertainties.
The obtained upper limits at 90\% C.L. are $\BR(\bzkxmu) < 2.01 \times 10^{-8}$ and $\BR(\bzrhoxmu) < 1.51 \times 10^{-8}$.

\section{Summary and Conclusion}
We searched for a scalar and vector particle in the decays $\bzkxmu$ and $\bzrhoxmu$. 
No significant signals are observed in a sample of $657 \times 10^6~B \bar B$ pairs. 
We set 90\% C.L. upper limits of $\BR(\bzkxmu) < 2.01 \times 10^{-8}$ 
and $\BR(\bzrhoxmu) < 1.51 \times 10^{-8}$; 
our results rule out models II and III for the sgoldstino interpretation of the HyperCP observation~\cite{demidov}.


\end{document}